\documentclass{iopart}
\usepackage{iopams}
\usepackage{cite}
\usepackage{graphicx}
\ifx\pdfoutput\undefined
\providecommand\ead[1]{\vspace*{5pt}\address{E-mail: \mailto{#1}}}
\def\mailto#1{{\texttt{\MakeLowercase{#1}}}}
\else
\usepackage{hyperref}
\hypersetup{%
  pdftitle={Bogomol'nyi decomposition for vesicles of arbitrary genus},
  pdfauthor={J\'er\^ome~Benoit, Avadh~Saxena and Turab~Lookman},
  pdfsubject={PACS: 02.40.-k, 87.16.Dg, 75.10.Hk, 11.27.+d},
  pdfpagemode=UseThumbs,
  pdfstartpage=1,
  pdfhighlight=/O,
  pdfview=FitH,
  pdfstartview=FitH,
  urlcolor=blue,
  }
\IfFileExists{thumbpdf.sty}{\usepackage{thumbpdf}}{}
\providecommand\ead[1]{\vspace*{5pt}\address{E-mail: \mailto{#1}}}
\def\mailto#1{\href{mailto:#1}{\texttt{\MakeLowercase{#1}}}}
\fi

\input{\jobname.rty}
\let\iint\int

\begin{document}
\paper[Bogomol'nyi decomposition for vesicles of arbitrary genus]{%
    Bogomol'nyi decomposition for vesicles\\ of arbitrary genus%
  $^\ast$%
  }
\author{%
  J\'er\^ome~Benoit$^{1,2}$,
  Avadh~Saxena$^{1}$\
  and Turab~Lookman$^{1}$%
  }
\address{$^{1}$\
  Theoretical Division, Los Alamos National Laboratory,
  Los Alamos,
  New Mexico 87545, USA %
  }
\address{$^{2}$\
  Laboratoire de Physique Th\'eorique et Mod\'elisation (\textsc{CNRS-ESA}~8089),
  Universit\'e de Cergy-Pontoise,
  5~Mail~Gay-Lussac,
  95031~Cergy-Pontoise,
  France%
  }
\address{$\ast$\
  \textnormal{Journal of Physics A: Mathematical and General} \textbf{34}, 9417 (2001) %
  [\texttt{cond-mat}/0103004]%
  }%
\ead{jgmbenoit@wanadoo.fr}
\date{\texttt{cond-mat}/0103004}
\begin{abstract}
We apply the Bogomol'nyi technique,
which is usually invoked in the study
of solitons or models with topological invariants,
to the case of elastic energy of vesicles.
We show that spontaneous bending contribution
caused by any deformation from metastable bending shapes
falls in two distinct topological sets:
shapes of spherical topology
and shapes of non-spherical topology
experience respectively
a deviatoric bending contribution \textit{\`a~la} Fischer
and a mean curvature bending contribution \textit{\`a~la} Helfrich.
In other words,
topology may be considered to describe bending phenomena.
Besides,
we calculate
the bending energy per genus and the bending closure energy
regardless of the shape of the vesicle.
As an illustration we briefly consider geometrical frustration phenomena
experienced by magnetically coated vesicles.
\end{abstract}


Our motivation is to amplify on
the observation of vesicles with arbitrary low genus
(number of holes/handles)
exhibiting conformal diffusion
(spontaneous conformal transformation),
namely existence of two conservation laws for vesicles
\cite{PLIPO,XAmMn,CFMV,FVNT,OSSCDGV,VTTOMST}.
The \textit{genus} is a topological invariant:
a quantity conserved under smooth transformations
which does not depend on the static or dynamic
equations of the system under consideration.
In contrast,
\textit{conformal diffusion} provides evidence that
the system Hamiltonian is invariant under
conformal transformations.
Whereas the N{\oe}ther theorem \cite{NOETHER,NOETHERTAVEL}
may be used to treat
the latter invariance law in order to compute
the corresponding conserved current and constant charge,
the Bogomol'nyi technique enables one to treat successfully
various models with topological invariants
\cite{BelavinPolyakov,Bogomolnyi,BogoE}.
In this Letter we focus on
the topological conservation law only;
we defer the conformal diffusion to future articles.

To obtain the Bogomol'nyi relationships
we write down
for vesicles of arbitrary genus
a bending Hamiltonian
as a covariant functional
invariant under conformal transformations
which depends on their shape only
and which is suitable for the Bogomol'nyi decomposition.
Applying the converse of
the \textit{remarkable theorem} of Gauss \cite{XBishopCrittenden}
enables us to construct such an Hamiltonian.
Instead of describing shapes by their \textit{Monge representation}
(\textit{i.e.}, their surface equation) as customary \cite{XAmMn,CFMV},
we characterize shapes
by their metric tensor and their shape tensor:
the total integral of the shape tensor self-product
is our successful candidate.
Then the Bogomol'nyi technique reveals
the topological nature of bending phenomena
while differential geometry extends
forward the Bogomol'nyi relationships.
In particular we show that
any deformation of the non-trivial metastable shapes spontaneously leads
to a deviatoric bending contribution \textit{\`a~la} Fischer
\cite{BSLBIII,BSLBV,NTSSCI}
for shapes of spherical topology
and to a mean curvature bending contribution
(up to a conformal transformation of the ambient space)
\textit{\`a~la} Helfrich
\cite{XAmMn,CFMV,HelfrichZN2}
for shapes of non-spherical topology:
our approach exhibits that
the bending contribution expression depends on the shape topology
---our main result concisely contained in formulae
(\ref{BF/Hb/decomposition/umbilical}),
(\ref{BF/Hb/decomposition/minimal})
and (\ref{BF/Hb/decomposition/closed}).

Before describing the bending energy of vesicles, we first succinctly
recall the Bogomol'nyi technique
[Eqs. (\ref{BogoTechnique/begin})--(\ref{BogoTechnique/end}) below]
through the nonlinear $\sigma$-model
\cite{BelavinPolyakov,Bogomolnyi,Felsager}.
More precisely
we consider spin fields on curved surfaces $\stSurface$
with the nonlinear $\sigma$-model as interaction
\cite{VSDTS,TSGF,HSETS,HSCS}:
\begin{equation}
\label{BogoTechnique/begin}
\label{SF/Hm/covariant}
\stHmMag=
  {\textstyle{\frac{1}{2}}}\stJ\!%
  {\iint\limits_\stSurface}\!\sqrt{g}\,{\std\Omega}\;%
  g^{ij}h_{\alpha\beta}\partial_{i}n^{\alpha}\partial_{j}n^{\beta}
  ,
\end{equation}
where the order parameter $n^{\alpha}$ corresponds to
a point on the two-sphere ${\mathbb{S}}^2$
and the phenomenological parameter $\stJ$ to
the coupling energy between neighbouring spins.
The metric tensors $g_{ij}$ and $h_{\alpha\beta}$ describe respectively
the support manifold $\stSurface$
(\textit{i.e.}, the underlying geometry)
and the order parameter manifold ${\mathbb{S}}^2$:
as customary, $g$ represents the determinant $\det(g_{ij})$
and $\sqrt{g}\,{\std\Omega}$ the area element.
Assuming homogeneous boundary conditions, when applicable,
allows one to map each boundary to a point,
then the support manifolds $\stSurface$ are topologically equivalent to
the torus ${\mathbb{T}}_\stgenus$ of genus $\stgenus$.
Consequently, the order parameter field $n^{\alpha}$ effects
the mapping of the compactified support ${\mathbb{T}}_\stgenus$
to the two-sphere ${\mathbb{S}}^2$ which is classified
by the cohomotopy group $\Pi^2({\mathbb{T}}_\stgenus)$ isomorphic to
the set of integers $\mathbb{Z}$ \cite{Shankar1977}:
two spin configurations belonging to distinct classes
cannot be smoothly deformed into one another.
Since there exists an infinite number of classes that
map the torus ${\mathbb{T}}_\stgenus$ of genus $\stgenus$ to the two-sphere ${\mathbb{S}}^2$,
for each compactified surface $\stSurface$
the space of spin configurations splits into
an infinite number of distinct components,
each characterized by a definite topological invariant.
Here the topological invariant is the degree of mapping $\stPontrjagin$
which is expressed in terms of the order parameter field $n^{\alpha}$ as
\begin{equation}
\label{SF/TI/covariant}
\stPontrjagin=
  {\textstyle{\frac{1}{8\pi}}}\!%
  {\iint\limits_\stSurface}\!\sqrt{g}\,{\std\Omega}\;%
  e^{ij}f_{\alpha\beta}\,\partial_{i}n^{\alpha}\partial_{j}n^{\beta}
  ,
\end{equation}
with $e_{ij}$ and $f_{\alpha\beta}$
the antisymmetric tensors associated
with the support manifold $\stSurface$ and
the target (i.e., the order parameter) manifold ${\mathbb{S}}^2$, respectively.
In spherical coordinates
the total integral (\ref{SF/TI/covariant}) immediately reads
as the winding number.
Note that the topological conservation law arises
from the nature of the order parameter field $n^{\alpha}$ only.
Henceforth we demonstrate how the Bogomol'nyi technique enables one
to study topological spin configurations
subject to the magnetic Hamiltonian (\ref{SF/Hm/covariant}).
By introducing the self-dual tensors
\begin{equation}
\label{SF/FT/def}
{T^{\pm}}_{i\alpha}\equiv
  {\textstyle{\frac{1}{\sqrt{2}}}}
  \left[
    \partial_{i}n_{\alpha}
    \mp\,
    e_{ir}f_{\alpha\kappa}\,
    \partial^{r}n^{\kappa}
  \right]%
,
\end{equation}
which satisfy the \textit{precious} inequalities
\cite{BelavinPolyakov,Bogomolnyi}
\begin{equation}
\label{SF/FT/inequalities}
{T^{\pm}}_{i\alpha}{T^{\pm}}^{i\alpha}
  \geqslant 0
,
\end{equation}
the magnetic Hamiltonian (\ref{SF/Hm/covariant}) density decomposes as
\begin{equation}
\label{SF/HmD/decomposition}
{\textstyle{\frac{1}{2}}}%
\partial_{i}n_{\alpha}\partial^{i}n^{\alpha}=
  {\textstyle{\frac{1}{2}}}%
  {T^{\pm}}_{i\alpha}{T^{\pm}}^{i\alpha}
  \pm\:%
  {\textstyle{\frac{1}{2}}}%
  e^{ij}f_{\alpha\beta}\,\partial_{i}n^{\alpha}\partial_{j}n^{\beta}
  .
\end{equation}
Integrating the previous decompositions (\ref{SF/HmD/decomposition})
over the support manifold $\stSurface$ and
inserting the formula (\ref{SF/TI/covariant})
lead us to rewrite the magnetic Hamiltonian (\ref{SF/Hm/covariant})
in each topological class specified
by the winding number $\stPontrjagin$ in the form
\begin{equation}
\label{SF/Hm/decomposition}
\stHmMag=
  {\textstyle{\frac{1}{2}}}\stJ\!%
  {\iint\limits_\stSurface}\!\sqrt{g}\,{\std\Omega}\;%
  {T^{\pm}}_{i\alpha}{T^{\pm}}^{i\alpha}
  +%
  4\pi\stJ\,\left|\stPontrjagin\right|
  ,
\end{equation}
with $(\pm)$ the sign of $\stPontrjagin$.
Readily from the \textit{precious} inequalities (\ref{SF/FT/inequalities}),
the magnetic Hamiltonian (\ref{SF/Hm/covariant}) yields
\begin{equation}
\label{SF/Hm/BogoBound}
\stHmMag\geqslant4\pi\stJ\,\left|\stPontrjagin\right|
  .
\end{equation}
Moreover,
in each topological class,
the lowest value of the magnetic energy is actually attained
when the self-dual tensor ${T^{\pm}}_{i\alpha}$ vanishes.
In other words, according to the definition (\ref{SF/FT/def}),
the metastable order parameter fields $n^{\alpha}$, which actually saturate
the Bogomol'nyi bound (\ref{SF/Hm/BogoBound}), satisfy
the first-order self-dual differential (or Bogomol'nyi) equation multiplet
\begin{equation}
\label{BogoTechnique/end}
\label{SF/Hm/SD/eq}
  \partial_{i}n^{\alpha}=
  \pm\:
  e_{i\hphantom{r}}^{\hphantom{i}r}
  f_{\hphantom{\alpha}\kappa}^{\alpha\hphantom{\kappa}}
  \partial_{r}n^{\kappa}
  .
\end{equation}
\newcounter{counterBogoFeature}\setcounter{counterBogoFeature}{1}%
Clearly the Bogomol'nyi decomposition
reveals the underlying topology:
(\roman{counterBogoFeature})\stepcounter{counterBogoFeature}~%
the inequality (\ref{SF/Hm/BogoBound}) claims that
the \textit{minimum minimorum} of the Hamiltonian
in each topological class is proportional
to the topological invariant;
(\roman{counterBogoFeature})\stepcounter{counterBogoFeature}~%
the metastable configurations which actually saturate the Hamiltonian
satisfy something simpler than the usual (Euler-Lagrange) equations
and no explicit solution is needed to compute their energy;
(\roman{counterBogoFeature})\stepcounter{counterBogoFeature}~%
from the computation emerges a self-dual symmetry;
(\roman{counterBogoFeature})\stepcounter{counterBogoFeature}~%
the decomposition (\ref{SF/Hm/decomposition}) clarifies
the stability of the topological configurations
and exhibits a spontaneous energy contribution
due to any deformation from metastable configurations;
(\roman{counterBogoFeature})\stepcounter{counterBogoFeature}~%
a deviatoric (or strain-like) tensor ${T^{\pm}}_{i\alpha}$ is constructed.
Usually not explored,
the last two features will appear very relevant below.

Now, we focus on vesicles of arbitrary genus.
The large separation of length scales
between the membrane thickness of vesicles and their overall size
allows us to describe each vesicle as a surface manifold $\stSurface$
embedded in the tridimensional space ${\mathbb{R}}^3$
\cite{XAmMn,CFMV}.
As a result, initially we focus only on
the energy induced by the surface manifold $\stSurface$ itself.
On the other hand, the corresponding Hamiltonian has to be
invariant under conformal transformations
and suitable for the Bogomol'nyi technique.
Nonetheless, we must first choose how to represent the surface
manifold $\stSurface$:
we invoke the converse of the \textit{remarkable theorem} of Gauss
\cite{XBishopCrittenden}.
Let $\stSurface(x^k)$ be a surface manifold
with arbitrary intrinsic coordinates $(x^k)$
embedded in the tridimensional ambient space ${\mathbb{A}}^3$,
then there uniquely exist
a metric tensor $g_{ij}(x^k)$ and a shape tensor $b_{ij}(x^k)$
related by a definite equation multiplet;
the converse is true except for the exact position of the surface
$\stSurface$ in the ambient space ${\mathbb{A}}^3$.
Then, with a different notation,
we concisely express the \textit{remarkable theorem} as
\begin{equation}
\label{RTG/theorem}
\stSurface(x^k)\subset{\mathbb{A}}^3
\;\Longleftrightarrow\;
\left(g_{ij}(x^k),b_{ij}(x^k)\right).
\end{equation}
The geometrical meaning of the couple $(g_{ij},b_{ij})$
in (\ref{RTG/theorem}) is revealed by introducing both
the infinitesimal tangential displacement $\stdT$
over the surface manifold $\stSurface$
and the infinitesimal normal displacement $\stdN$
to the surface manifold $\stSurface$
inside the ambient space ${\mathbb{A}}^3$;
we have
\numparts
\label{RTG/form}
\begin{eqnarray}
\label{RTG/form/first}
\stdT\!\cdot\!\stdT&=g_{ij}(x^k)\,\std{x^{i}}\std{x^{j}},\\
\label{RTG/form/second}
\stdN\!\cdot\!\stdT&=b_{ij}(x^k)\,\std{x^{i}}\std{x^{j}}.
\end{eqnarray}
\endnumparts
In differential geometry, the quadratic differential forms
(\ref{RTG/form/first}) and (\ref{RTG/form/second}) are referred to as
the \textit{first} and \textit{second} \textit{fundamental forms},
respectively.
Furthermore, the Riemann tensor,
which measures how much a manifold is curved,
reduces to
\begin{equation}
\label{RTG/curvature/Riemann}
R_{klmn}=%
  \stGC\:%
  \left[g_{km}g_{ln}-g_{kn}g_{lm}\right],
\end{equation}
where
the Gaussian curvature $\stGC$ depends only on the metric tensor $(g_{ij})$
and its first and second derivatives:
such an entity is called a \textit{bending invariant}
or is said to be \textit{intrinsic} \cite{XBishopCrittenden,Struik}.
When the surface manifold $\stSurface$ is embedded
in a curved tridimensional ambient space ${\mathbb{A}}^3$,
then the intrinsic curvature
$\stGC$ splits as
\begin{equation}
\label{RTG/curvature/decomposition}
\stGC=\stEC+\stSC ,
\end{equation}
where the \textit{extrinsic} curvature $\stEC$ satisfies
\begin{equation}
\label{RTG/curvature/extrinsic}
\stEC=%
  {\textstyle{\frac{1}{2}}}\:%
  e^{ij}e^{kl}\,b_{ik}b_{jl} ,
\end{equation}
whereas the \textit{sectional} curvature $\stSC$ characterizes the
ambient space.
For example,
the sectional curvature $\stSC$ vanishes
for the flat space ${\mathbb{R}}^3$
and is equal to one
for the three-sphere ${\mathbb{S}}^3$.
We complete this brief overview by
stating the Gauss-Bonnet theorem \cite{Struik,CEMSFTC}:
\begin{equation}
\label{BF/TI/GaussBonnetChern}
{\iint\limits_{{\stSurface_{\stgenus,\stends}}\subset{\mathbb{A}}^3}}\!%
  \sqrt{g}\,{\std\Omega}\:%
  \stGC%
  =%
  -4\pi\left(\stgenus\!+\!\stends\!-\!1\right)%
\end{equation}
with $\stSurface_{\stgenus,\stends}$
a surface manifold topologically equivalent to a closed surface manifold
of genus $\stgenus$ less $\stends$ points
(ends)
embedded in ${\mathbb{A}}^3$.

With this background, we now assume the bending Hamiltonian as the
following functional
\begin{equation}
\label{BF/Hb/covariant}
\stHmBnd\equiv%
  {\textstyle{\frac{1}{2}}}\stKb\!%
  {\iint\limits_{\stSurface\subset{\mathbb{R}}^3}}\!%
  \sqrt{g}\,{\std\Omega}\:%
  g^{ij}g_{kl}\,b_{i}^{\hphantom{i}k}b_{j}^{\hphantom{j}l}%
  ,
\end{equation}
where the phenomenological parameter $\stKb$
describes the bending rigidity.
Since the total integral in (\ref{BF/Hb/covariant})
is known to be invariant under conformal transformations \cite{AICM,OPCW},
it remains to show that our assumption
fits the Bogomol'nyi technique as desired.
To begin, note that the bending Hamiltonian $\stHmBnd$ formula
(\ref{BF/Hb/covariant}) stresses the similitude with the magnetic
Hamiltonian $\stHmMag$ (\ref{SF/Hm/covariant}).  Then the self-dual
deviatoric tensors are defined as
\begin{equation}
\label{BF/FT/def}
{B^{\pm}_{\hphantom{\pm}}}_{ij}\equiv%
  {\textstyle{\frac{1}{\sqrt{2}}}}
  \left[
    b_{ij}\mp\,e_{ik}e_{jl}\:b^{kl}%
  \right]
\end{equation}
which yield the \textit{precious} relationships
\begin{equation}
\label{BF/FT/relationships}
{B^{\pm}_{\hphantom{\pm}}}_{ij}{B^{\pm}_{\hphantom{\pm}}}^{ij}=%
  \left(
    \stPC\mp\stpC
  \right)^2
  \geqslant0
\end{equation}
where $\stPC$ and $\stpC$ denote
the eigenvalues of the shape tensor $(b^{i}_{\hphantom{i}j})$,
namely the principal curvatures of the surface (or support) manifold $\stSurface$.
Here the decompositions (\ref{SF/HmD/decomposition}) read
\begin{equation}
\label{BF/FT/decomposition}
{\textstyle{\frac{1}{2}}}%
b_{ij}b^{ij}=%
  {\textstyle{\frac{1}{2}}}%
  {B^{\pm}_{\hphantom{\pm}}}_{ij}{B^{\pm}_{\hphantom{\pm}}}^{ij}%
  \pm\,\stEC%
  .
\end{equation}
By inserting the previous decomposition (\ref{BF/FT/decomposition})
in the formula (\ref{BF/Hb/covariant})
and recognizing the total curvature (\ref{BF/TI/GaussBonnetChern}),
the bending Hamiltonian $\stHmBnd$ (\ref{BF/Hb/covariant}) decomposes as
\begin{equation}
\label{BF/Hb/decomposition/rough}
\stHm_{b}%
  =%
  {\textstyle{\frac{1}{2}}}\stKb\!%
  {\iint\limits_{{\stSurface_{\stgenus,\stends}}\subset{\mathbb{R}}^3}}\!%
  \sqrt{g}\,{\std\Omega}\:%
  {B^{\pm}_{\hphantom{\pm}}}_{ij}{B^{\pm}_{\hphantom{\pm}}}^{ij}%
  \mp%
  4\pi\stKb\left(\stgenus\!+\!\stends\!-\!1\right)%
  .
\end{equation}
Note that we have used the fact that,
according to (\ref{RTG/curvature/decomposition}),
the extrinsic curvature $\stEC$
and the intrinsic curvature $\stGC$ are equal
when the ambient space is flat $(\stSC=0)$.
Existence theorems select the correct value for the sign $(\pm)$
and the proper topological classes for vesicles as follows.
From the \textit{precious} relationships (\ref{BF/FT/relationships}) it is
evident that the surface manifolds $\stSurface$ which actually saturate
the Bogomol'nyi decomposition (\ref{BF/Hb/decomposition/rough}) are
the \textit{totally umbilical surfaces} $(\stPC=\stpC)$
and the \textit{minimal surfaces} $(\stPC+\stpC=0)$.
Since only the round two-sphere ${\mathbb{S}}^2$ is totally umbilical \cite{Struik,CEMSFTC},
the decomposition with sign $(+)$ is relevant
only for surface manifolds $\stSurface_{0}$
topologically equivalent to the round two-sphere ${\mathbb{S}}^2$;
our first key result reads
\begin{equation}
\label{BF/Hb/decomposition/umbilical}
\stHm_{b}\left[{\stSurface_{0}}\right]
  =%
  {\textstyle{\frac{1}{2}}}\stKb\!%
  {\iint\limits_{{\stSurface_{0}}\subset{\mathbb{R}}^3}}\!%
  \sqrt{g}\,{\std\Omega}\:%
  {B^{+}_{\hphantom{+}}}_{ij}{B^{+}_{\hphantom{+}}}^{ij}%
  +%
  4\pi\stKb%
  .
\end{equation}
It is noticeable
that any deformation of the metastable bending configurations
for shapes of spherical topology
spontaneously leads to
a deviatoric bending contribution
\textit{\`a~la} Fischer \cite{BSLBIII,BSLBV,NTSSCI}:
to the best of our knowledge,
there is no direct derivation in the literature
for such a bending contribution suggested first by Fischer \cite{BSLBIII}.
Since within the flat ambient space ${\mathbb{R}}^3$
there is no closed minimal surface $(\stends=0)$
whereas minimal surfaces of genus $\stgenus$ $(\stgenus\geqslant0)$
with $\stends$ ends $(\stends\geqslant2)$ do exist \cite{CEMSFTC},
the decomposition with sign $(-)$ is relevant only for surface manifolds
$\stSurface_{\stgenus,\stends}$ topologically equivalent to such minimal
surfaces $\stMinSurface_{\stgenus,\stends}$;
our second key result immediately reads
\begin{equation}
\label{BF/Hb/decomposition/minimal}
\stHm_{b}\left[{\stSurface_{\stgenus,\stends}}\right]
  =%
  {\textstyle{\frac{1}{2}}}\stKb\!\!%
  {\iint\limits_{{\stSurface_{\stgenus,\stends}}\subset{\mathbb{R}}^3}}\!\!%
  \sqrt{g}\,{\std\Omega}\:%
  {B^{-}_{\hphantom{-}}}_{ij}{B^{-}_{\hphantom{-}}}^{ij}%
  +%
  4\pi\stKb\left(\stgenus\!+\!\stends\!-\!1\right)%
  .
\end{equation}
Here any deformation of the metastable bending configurations
spontaneously leads to a mean curvature
(or anti-deviatoric)
bending contribution \textit{\`a~la} Helfrich \cite{HelfrichZN2}
widely employed to describe bending phenomena \cite{PLIPO,XAmMn,CFMV}.
Note that the minimal surface $\stMinSurface_{0,2}$,
namely the \textit{catenoid},
is the elementary neck used to build or analyze numerically
vesicles of arbitrary genus $\stgenus$ \cite{XAmMn,CFMV,FVNT}.
We cannot proceed much further unless we invoke
an assertion due to Chen which claims \cite{AICM,OPCW}
\begin{equation}
\label{BF/Hb/Chen}
  {\iint\limits_{\stSurface\subset{\mathbb{R}}^3}}\!%
  \sqrt{g}\,{\std\Omega}\:%
  {\textstyle{\frac{1}{2}}}\,%
  b_{ij}b^{ij}%
  =\!\!%
  {\iint\limits_{\stConf(\stSurface)\subset\stConf({\mathbb{R}}^3)}}\!\!\!\!\!\!\!%
  \sqrt{g^{\stConf}}\,{\std\Omega^{\stConf}}\!%
  \left[
    {\textstyle{\frac{1}{2}}}\,%
    {b^{\stConf}}_{ij}{b^{\stConf}}^{ij}%
    \!+\!%
    \stSC^{\stConf}%
  \right]
  ,
\end{equation}
where $\stConf$ corresponds to an arbitrary conformal transformation.
Consequently the bending Hamiltonian $\stHmBnd$ (\ref{BF/Hb/covariant})
extends as
\begin{equation}
\label{BF/Hb/extended}
\stHmBnd=
  {\textstyle{\frac{1}{2}}}\stKb\!\!\!\!\!\!%
  {\iint\limits_{\stConf(\stSurface)\subset\stConf({\mathbb{R}}^3)}}\!\!\!\!\!\!%
  \sqrt{g^{\stConf}}\,{\std\Omega^{\stConf}}\!%
  \left[
    {b^{\stConf}}_{ij}{b^{\stConf}}^{ij}%
    \!+\!%
    2\stSC^{\stConf}%
  \right]
  .
\end{equation}
It is easily seen that the extended bending Hamiltonian
$\stHmBnd$ (\ref{BF/Hb/extended}) decomposes as
\begin{eqnarray}
\label{BF/Hb/decomposition/extended}
\stHm_{b}%
  \!=%
  {\textstyle{\frac{1}{2}}}\stKb\!\!\!\!\!\!\!\!%
  {\iint\limits_{\stConf(\stSurface_{\stgenus,\stends})\subset\stConf({\mathbb{R}}^3)}}\!\!\!\!\!\!\!\!\!%
  \sqrt{g^{\stConf}}\,{\std\Omega^{\stConf}}\!%
  {{B^{\stConf}}^{\pm}_{\hphantom{\pm}}}_{ij}{{B^{\stConf}}^{\pm}_{\hphantom{\pm}}}^{ij}
  \mp%
  4\pi\stKb\left(\stgenus\!+\!\stends\!-\!1\right)%
  +%
  2\stKb{\stClosure^{\pm}_\stConf}%
  \nonumber\\
  \hphantom{\stHm_{b}\!={\textstyle{\frac{1}{2}}}\stKb\iint}%
  \textrm{with}\quad%
  \stClosure^{+}_\stConf\!=\!0%
  \quad\textrm{and}\quad%
  \stClosure^{-}_\stConf\!=\!\!\!\!\!\!\!%
    {\iint\limits_{\stConf(\stSurface_{\stgenus,\stends})\subset\stConf({\mathbb{R}}^3)}}\!\!\!\!\!\!\!\!%
    \sqrt{g^{\stConf}}\,{\std\Omega^{\stConf}}\:%
    \stSC^{\stConf}%
  .
\end{eqnarray}
Consequently, the Chen assertion (\ref{BF/Hb/Chen}) strongly suggests
to apply existence theorems for minimal surfaces
living in curved tridimensional space ${\mathbb{A}}^3$.
To illustrate this point,
we will establish Bogomol'nyi relationships for closed surfaces.
A theorem due to Lawson \cite{CMSS3} claims that there exist
closed minimal surfaces $\stLawsonXi_{m,n}$ of arbitrary genus $\stgenus=m\,n$
in the three-sphere ${\mathbb{S}}^3$ $(\stSC=1)$;
$\stLawsonXi_{0,n}$ is the round two-sphere ${\mathbb{S}}^2$,
$\stLawsonXi_{1,1}$ the flat torus (\textit{i.e.}, the Clifford torus).
Furthermore,
the Willmore-Kusner conjecture \cite{Willmore,WillmoreRG,GGESTS,CSWP} asserts that
the surfaces $\stLawsonXi_{\stgenus,1}$ actually minimize the functional
$\stClosure^{-}_\stConf$ in (\ref{BF/Hb/decomposition/extended}):
we may have
\begin{equation}
\label{BF/Hb/WillmoreKusner}
\stWillmoreKusner_{\stgenus}\equiv%
  \inf_{\stSurface_{\stgenus,0}\subset{\mathbb{R}}^3}%
  \left[
    \hphantom{x}%
    {\iint\limits_{\stConf(\stSurface_{\stgenus,0})\subset\stConf({\mathbb{R}}^3)}}\!\!\!\!\!\!\!\!%
    \sqrt{g^{\stConf}}\,{\std\Omega^{\stConf}}\:%
    \stSC^{\stConf}%
  \right]%
  =\!%
  {\iint\limits_{\stLawsonXi_{\stgenus,1}\subset{{\mathbb{S}}^3}}}\!\!\!%
  \sqrt{g}\,{\std\Omega}%
  .
\end{equation}
Readily the Bogomol'nyi decomposition for closed surfaces
$(\stends=0)$ takes the form
\begin{equation}
\label{BF/Hb/decomposition/closed}
\stHm_{b}\left[{\stSurface_{\stgenus}}\right]
  \!=\!%
  {\textstyle{\frac{1}{2}}}\stKb\!\!\!\!\!%
  {\iint\limits_{\stConf(\stSurface_{\stgenus})\subset{{\mathbb{S}}^3}}}\!\!\!\!\!\!%
  \sqrt{g^{\stConf}}\,{\std\Omega^{\stConf}}\!%
  {{B^{\stConf}}^{-}_{\hphantom{-}}}_{ij}{{B^{\stConf}}^{-}_{\hphantom{-}}}^{ij}
  +%
  4\pi\stKb\left(\stgenus\!-\!1\right)%
  +%
  2\stKb\stWillmoreKusner_{\stgenus} ,%
\end{equation}
where the conformal transformation $\stConf$ maps ${\mathbb{R}}^3$ to ${\mathbb{S}}^3$.
Accordingly,
the Bogomol'nyi technique combined with the Chen assertion (\ref{BF/Hb/Chen})
allows us to measure how much a surface manifold is deformed from any surface
which is minimal inside a certain curved ambient space ${\mathbb{A}}^3$.
Thus, without loss of generality,
the Willmore-Kusner conjecture (\ref{BF/Hb/WillmoreKusner})
enables us to outline
through the formula (\ref{BF/Hb/decomposition/closed})
our third key result:
any deformation of the metastable bending configurations
for shapes of non-spherical topology
spontaneously leads to a mean curvature bending contribution
\textit{\`a~la} Helfrich
up to a conformal transformation of the ambient space.
\begin{figure}[t]
  \begin{center}
    \includegraphics[width=.7071\linewidth]{bdvag-LawsonXi_genus.mps}%
  \end{center}
  \caption{%
    Lawson sequence $\stWillmoreKusner_{\stgenus}$:
    \textsl{circles} represent exact values,
    \textsl{crosses} numerical rough estimates computed
    with Brakke's \texttt{Surface Evolver} \protect\cite{SEvlv}.
    The bold fitted lines describe the estimate of $\stWillmoreKusner_{\stgenus}$
    as $\stgenus$ tends to infinity \protect\cite{GGESTS}:
    $\stWillmoreKusner_{\stgenus}=8\pi\!-\!{\mathsf{c}}/{\stgenus}\!+\!O({1}/{\stgenus^2})$
    where $-\mathsf{c}$ is the slope of the inset plot.
    }%
\label{fig/LawsonXi_genus}
\end{figure}
Before summarizing
let us expose how the bending energy bounds
(\textit{i.e.}, the Bogomol'nyi bounds)
are governed.
From the Bogomol'nyi decompositions (\ref{BF/Hb/decomposition/umbilical}),
(\ref{BF/Hb/decomposition/minimal}) and (\ref{BF/Hb/decomposition/closed})
the Bogomol'nyi bounds are:
\begin{equation}
\label{BF/Hb/BogoBound}
\stHm_{b}\!\left[{\stSurface_{\stgenus,\stends}}\right]
  \!\geqslant\!%
  \left\{
    \begin{array}{ll}
      2\stWillmoreKusner_{\stgenus}\stKb\!+\!4\pi\stKb\left(\stgenus\!-\!1\right)&
        \textrm{if $\stgenus\!\geqslant\!0$ and $\stends\!=\!0$},\\
      4\pi\stKb\left(\stgenus\!+\!\stends\!-\!1\right)&
        \textrm{if $\stgenus\!\geqslant\!0$ and $\stends\!\geqslant\!2$}.\\
    \end{array}
  \right.
\end{equation}
Obviously the Bogomol'nyi bounds for vesicles (\ref{BF/Hb/BogoBound})
are proportional to the bending rigidity $\stKb$ and separate into two parts:
a closure bending energy $2\stWillmoreKusner_{\stgenus}\stKb$ and
a genus/end bending energy $4\pi\stKb\left(\stgenus\!+\!\stends\!-\!1\right)$.
Whereas the bending energy per genus/end
is straightforward to compute $(4\pi\stKb)$,
there exists no literal formula for the Lawson sequence
$\stWillmoreKusner_{\stgenus}$ yet:
for the round two-sphere ${\mathbb{S}}^2$ and the flat torus $\stLawsonXi_{1,1}$
the Lawson sequence $\stWillmoreKusner_{\stgenus}$ takes respectively as exact value
$\stWillmoreKusner_{0}\!=\!4\pi$ and $\stWillmoreKusner_{1}\!=\!2\pi^2$
(\textsl{circles} in Fig.~\ref{fig/LawsonXi_genus}),
for higher genus we have computed numerical estimates
(\textsl{crosses} in Fig.~\ref{fig/LawsonXi_genus}).
\newcounter{counterBogoDiffGeoSumUp}\setcounter{counterBogoDiffGeoSumUp}{1}%

To summarize our key results concisely expressed in formulae
(\ref{BF/Hb/decomposition/umbilical}),
(\ref{BF/Hb/decomposition/minimal})
and (\ref{BF/Hb/decomposition/closed})
the Bogomol'nyi technique, extended forward
with existence theorems from differential geometry,
allows us to show that
any deformation of the non-trivial metastable shapes
(which are clearly identified)
generates a bending elastic energy contribution
which falls in two distinct categories
with respect to the topology of the shape:
(\roman{counterBogoDiffGeoSumUp})\stepcounter{counterBogoDiffGeoSumUp}~%
for shapes of spherical topology
a deviatoric bending contribution \textit{\`a~la} Fischer
\cite{BSLBIII,BSLBV,NTSSCI}
spontaneously arises;
(\roman{counterBogoDiffGeoSumUp})\stepcounter{counterBogoDiffGeoSumUp}~%
for shapes of non-spherical topology
a mean curvature bending contribution \textit{\`a~la} Helfrich
\cite{XAmMn,CFMV,HelfrichZN2}
(up to a conformal transformation of the ambient space)
spontaneously emerges.
We observe that this splitting comes from
the breaking of the self-dual symmetry
of the Bogomol'nyi relationships by
existence theorems from differential geometry.
Besides,
the Bogomol'nyi relationships
allow us to compute both
the bending energy per genus/end
and the bending closure energy
for vesicles regardless of their shape.

As a further illustration,
our approach leads to a clear understanding of
geometrical frustration phenomena
experienced by magnetically coated vesicles \cite{VSDTS,TSGF,HSETS,HSCS}:
the presence of the double Bogomol'nyi decomposition
generates a competition between magnetic solitons and shapes which
tend to saturate the magnetic energy (\ref{SF/Hm/covariant})
and the bending energy (\ref{BF/Hb/covariant}), respectively.
More precisely,
when at least one deviatoric tensor cannot vanish,
the balance between the two deviatoric energies releases the frustration:
hence both magnetic and geometric effects manifest
in accordance with the deviatoric energies,
\textit{e.g.} by removing a mismatch between magnetic and geometric
(or underlying support) length scales.

In conclusion,
our approach gives in a rather natural manner
the Bogomol'nyi relationships for vesicles:
their typical features
combined with existence theorems from differential geometry
show that spontaneous bending deformation
from metastable bending shapes
splits in two distinct topological classes
(shapes of spherical topology and shapes of non-spherical topology):
in other words, topology may be considered to describe bending
phenomena--in contradiction
with customary phenomenological approaches \cite{XAmMn,CFMV}.
Furthermore,
the appearance of Bogomol'nyi relationships for vesicles
provides a powerful guide for understanding vesicles
and enlarges the application field of
the Bogomol'nyi technique
(traditionally used in fields ranging
from condensed matter physics to high energy physics)
to elastic and geometrical phenomena in soft condensed matter.

This work was supported in part
by the U.S. Department of Energy.

\section*{References}
\bibliographystyle{jphya}
\bibliography{bdvag}

\end{document}